\documentclass[preprint,showpacs,showkeys,preprintnumbers,amsmath,amssymb]
{revtex4}

\usepackage{graphicx}
\usepackage{dcolumn}
\usepackage{bm}
\usepackage[latin1]{inputenc}

\begin{document}

\author{R. Rossi Jr.}
\author{M. C. Nemes}
\affiliation{Departamento de F\'{\i}sica, Instituto de Ciências
Exatas, Universidade Federal de Minas Gerais, C.P. 702, 30161-970,
Belo Horizonte, MG, Brazil}
\author{A. R. Bosco de Magalh\~{a}es}
\affiliation{Departamento Acadêmico de Ciências Básicas, Centro
Federal de Educação Tecnológica de Minas Gerais, 30510-000, Belo
Horizonte, MG, Brazil}

\title{Quantum Zeno Effect in Cavity QED: Experimental Proposal with Non
Ideal Cavities and Detectors}

\begin{abstract}
We propose an experiment with two coupled microwave cavities and a
``tunneling'' photon observed by the passage of Rydberg atoms. We
model the coupled cavities as in Ref. \cite{art1} and include
dissipative effects as well as limited detection efficiency. We also
consider realistic finite atom-field interaction times and provide
for a simple analytical expression for the photon ``tunneling''
probability including all these effects. We show that for
sufficiently small dissipation constants the effect can be observed
with current experimental facilities.
\end{abstract}

\maketitle

\section{Introduction}

The success of physical theories is intimately connected to its
potentiality to describe existing empirical data and to predict new,
yet to be observed, phenomena \cite{art2}. However the
interpretation of empirical data is not completely independent of
the proposed theory. Therefore in natural sciences the measurement
process plays a double role: it is at the same time a testing tool
of theories and also a physical process in itself, subjected to
theoretical analysis. In the quantum domain theoretical descriptions
of the measurement process are a matter of innumerous discussions.

In 1932, in his famous treatise \cite{art3}, J. von Neumann proposed
a quantum measurement theory, which became quickly well known. An
initial premise of this theory is the postulate that the measurement
of a given observable always yields one of the eigenvalues of this
observable and, after the measurement, the system collapses to the
corresponding eigenvector. This working hypothesis is known as
``projection postulate'' and is responsible for several
counterintuitive aspects of the theory. It has led to the
formulation of several paradoxes.

The ``Quantum Zeno Paradox'' was presented in a mathematically
rigorous fashion in 1977 by B. Misra and E. C. Sudarshan
\cite{art4}. In this formulation the authors show that a sequence of
projective measurements on a system inhibits its time evolution. The
paradoxical character of this conclusion becomes explicit when one
continuously observes the state of an unstable particle. When the
Quantum Zeno Effect (QZE) was first formulated, it has been
associated to two factors: an initially quadratic time decay and the
projection postulate.

In the 90ies, after the realization of the pioneer experiment
\cite{art5} on the effect, which showed the interruption of the time
evolution of a decaying system by means of continuous observations,
the QZE became the center of fervorous debates \cite{art6,art7}. The
role attributed to the projection postulate was at the center of the
discussions. New approaches have been proposed \cite{art6,art8} and
the strong association between the QZE and the projection postulate
was no longer a necessary ingredient. Nowadays the literature on the
subject is vast and range from experimental proposals to fundamental
theoretical questions \cite{art9,art10,art11,art12}.

In the present contribution we will study the dynamics of the QZE in
a (apparently feasible \cite{art1,art13,art14}) experiment involving
two coupled microwave cavities, one photon and Rydberg atoms as
probes. The novel aspects explored here are the effect of a lossy
environment and of limited efficiency detection on the visibility of
the QZE.

In Section II, we describe the main elements of the proposed
experiment and their interaction. In Section III, the QZE is
investigated in the situation where several atoms interact with one
cavity mode and next with ionization detectors. In Section IV, we
show that these measurements of several atomic states are not
essential for the QZE; the effects of finite atom-field interaction
times and of field dissipation are also studied in this section. In
section V we draw the conclusions.

\section{The model for an Experiment}

Let us consider two cavity modes coupled by a conducting wire (wave
guide), as proposed in \cite{art1}. The Hamiltonian for the system
is given by
\begin{equation}
H_{AB}=\hbar \omega a^{\dagger }a+\hbar \omega b^{\dagger }b+\hbar
g(a^{\dagger }b+b^{\dagger }a),  \label{Hamiltonian}
\end{equation}
where $a^{\dagger }$ ($a$) and $b^{\dagger }$ ($b$) are creation
(annihilation) operators for modes $M_{A}$ and $M_{B}$, $\omega $
their frequency and $g$ a coupling constant \cite{art1}. The
situation we shall consider concerning the electromagnetic degree of
freedom will always involve the following initial state
\begin{equation*}
\rho _{F}(0)=|1_{A},0_{B}\rangle \langle 1_{A},0_{B}|=|1,0\rangle
\langle 1,0|,
\end{equation*}
where the bra (ket) $\left| n,m\right\rangle $ ($\left\langle
n,m\right| $)
refers to $n$ excitations in mode $M_{A}$ and $m$ excitations in mode $M_{B}$%
. The evolution of this state according to (\ref{Hamiltonian}) in a
time interval $T$ is given by
\begin{equation}
\rho _{F}(T)=|c_{1}(T)|^{2}|1,0\rangle \langle
1,0|+|c_{2}(T)|^{2}|0,1\rangle \langle 0,1|+(c_{1}(T)c_{2}^{\ast
}(T)|1,0\rangle \langle 0,1|+h.c.),
\end{equation}
where $c_{1}(T)=\cos (gT)$, $c_{2}(T)=\sin (gT)$ and $h.c.$ stands
for Hermitian conjugate. Thus, due to the coupling between the
cavities, a photon initially in cavity $A$ may be found at time $T$
in cavity $B$ with probability $|c_{2}(T)|^{2}$. At $T=\pi /2g$ the
photon has performed a complete transition from mode $M_{A}$ to mode
$M_{B}$: $\rho _{C}(T)=|0,1\rangle \langle 0,1|$.

In order to experimentally verify the occurrence of this transition,
one can measure the number of photons in cavity $B$: if the value
found is zero we know for sure that the transition did not occur.
This may be realized by sending an effectively two level atom
\cite{art15} in its lowest state through cavity $B$. The atom
prepared in its lowest state works as a probe for the field state.
In order to realize this ``two level atom'' one uses a
Rydberg atom whose relevant transition may be tuned to the field \emph{quanta%
} $\hbar \omega $. We denote by $|e\rangle $ ($|g\rangle $) the
higher (lower) energy atomic states. This tuning may be effected by
using a quadratic Stark effect, as in Ref. \cite{art16}. The control
of the atom-field interaction time may be performed by this method
with a precision of 1$\mu s$. Since this time is small compared to
the other relevant times in the experiment, we will not consider
imperfections in the atom-field
interaction time. The interaction of the atom with the field mode in cavity $%
B$ may be described by the Jaynes-Cummings model, which gives $\tau
_{\pi }=\pi /\Omega _{0}$, where $\Omega _{0}$ is the vacuum Rabi
frequency, for
the $\pi $ pulse time, the time in which one excitation moves from mode $%
M_{B}$ to the atom. If the atom-field coupling is much stronger than
the coupling between modes $M_{A}$ and $M_{B}$, $\tau _{\pi }$ may
be disregarded \footnote{In Section (IV) we will consider the
effects of $\tau _{\pi }$, taking its value from experimental
data.}, and we may write the density operator for the system
composed by the atom an the field modes, after the atom-field
interaction, as
\begin{equation}
\rho _{AF}(T)=|c_{1}(T)|^{2}|1,0,g\rangle \langle
1,0,g|+|c_{2}(T)|^{2}|0,0,e\rangle \langle
0,0,e|+(c_{1}(T)c_{2}^{\ast }(T)|1,0,g\rangle \langle 0,0,e|+h.c.).
\label{rhoca1}
\end{equation}
Since the atom-field state is maximally entangled, to measure the
atomic level in an ionization detector is equivalent to measuring
the number of photons in each cavity before the atom-field
interaction.

\section{The Detection Process}

In this section we will consider the measurement of the atomic state
by ionization detectors $D_{e}$ and $D_{g}$ constructed in such a
way as to ionize the atom in states $|e\rangle $ and $|g\rangle $
respectively.

\subsection{Perfect Detectors}

If one has perfect detectors, each atom sent through cavity $B$ will
produce a click either in $D_{e}$ or $D_{g}$. Thus the probability
$p_{1,0}$ that a photon initially in mode $M_{A}$ did not reach
cavity $B$ is equal to the probability $p_{clickD_{g}}$ of one click
in detector $D_{g}$ :
\begin{equation}
p_{1,0}=p_{clickD_{g}}=|c_{1}(T)|^{2}.
\end{equation}

If we send $N$ atoms, one at each time $t=iT_{0}/N$ ($i=1$ to $N$),
during the fixed time interval $T_{0}=\pi /2g$, we can in principle
monitor the photon transition from mode $M_{A}$ to mode $M_{B}$. The
temporal evolution of the system under such conditions consists of
$N$ steps composed by a free evolution during a time interval $\tau
_{A,B}=T_{0}/N$, followed by an atom-field interaction.

If in one of these steps we observe one click in $D_{e}$, we must
conclude
that the photon was found in cavity $B$. As may be seen in Eq. (\ref{rhoca1}%
), after this click the field state becomes $\rho _{F}=|0,0\rangle
\langle 0,0|$, and all the subsequent atoms will be detected in
$|g\rangle $ state.

Let us now consider an experimental sequence where no clicks in
$D_{e}$ are observed. At time $t=0$, the state of the atom-field
system is given by
\begin{equation}
\rho _{AF}(0)=|1,0,g\rangle \langle 1,0,g|,
\end{equation}
and during the period $\tau _{A,B}=T_{0}/N$ the system evolves under
the Hamiltonian (\ref{Hamiltonian}):
\begin{equation}
\rho _{AF}(\tau _{A,B})=|c_{1}(\tau _{AB})|^{2}|1,0,g\rangle \langle
1,0,g|+|c_{2}(\tau _{AB})|^{2}|0,1,g\rangle \langle
0,1,g|+(c_{1}(\tau _{AB})c_{2}^{\ast }(\tau _{AB})|1,0,g\rangle
\langle 0,1,g|+h.c.).
\end{equation}
At time $\tau _{A,B}$, the atom and the mode $M_{B}$ perform a $\pi
$ pulse (regarded as instantaneous), what leads to
\begin{equation}
\bar{\rho}_{AF}(\tau _{A,B})=|c_{1}(\tau _{AB})|^{2}|1,0,g\rangle
\langle 1,0,g|+|c_{2}(\tau _{AB})|^{2}|0,0,e\rangle \langle
0,0,e|+(c_{1}(\tau _{AB})c_{2}^{\ast }(\tau _{AB})|1,0,g\rangle
\langle 0,0,e|+h.c.).
\end{equation}
If we observe a click in $D_{g}$, the state of the system ends up in
\begin{equation}
\rho _{AF}(\tau _{AB})=|1,0,g\rangle \langle 1,0,g|=\rho _{AF}(0).
\end{equation}
The probability of such a click in the first step is $|c_{1}(\tau
_{AB})|^{2} $, and in this case the system is reprepared in state $%
|1,0,g\rangle \langle 1,0,g|$. If all atoms are detected in
$|g\rangle $ state, the evolution will be composed by $N$
\emph{identical steps} to the one just described. Thus the
probability of $N$ clicks in $D_{g}$ is
\begin{equation}
p_{clickD_{g}}^{(N)}=\left( |c_{1}(\tau _{AB})|^{2}\right) ^{N},
\end{equation}
which is equal to the probability $p_{1,0}^{(N)}$ that the photon is
still in cavity $A$ at time $T_{0}$, after the interaction between
the field and the $N$ atoms. If we consider the limit $N\rightarrow
\infty $,
\begin{equation}
\lim_{N\rightarrow \infty }p_{1,0}^{(N)}=\lim_{N\rightarrow \infty
}p_{clickD_{g}}^{(N)}=1.
\end{equation}
Zeno effect becomes explicit: the continuous measuring of the number
of photons in cavity $B$ inhibits the transition of the photon from
cavity $A$ to cavity $B$.

\subsection{Inefficient Detectors}

In order to take the limited efficiency of the detectors into
account we need a model for the detection process. In what follows
we consider a schematic model for the atom-detector interaction
\cite{art17}:
\begin{eqnarray}
H_{D} &=&\hbar \epsilon _{g}|g\rangle \langle g|+\hbar \epsilon
_{e}|e\rangle \langle e|+\hbar \int dk\epsilon _{k}|k\rangle \langle
k|
\notag \\
&&+\hbar v_{g}\int dk(|g\rangle \langle k|+|k\rangle \langle
g|)+\hbar v_{e}\int dk(|e\rangle \langle k|+|k\rangle \langle e|),
\label{HamiltonianDgDe}
\end{eqnarray}
where $|e\rangle $ and $|g\rangle $ represent the same atomic levels
as in previous sections, and the set $\left\{ |k\rangle \right\} $
concerns the continuum of atomic levels related to the ionization of
the atom. We next consider several possibilities.

\subsubsection{Only Detector $D_{g}$ is Present}

This case corresponds to the Hamiltonian (\ref{HamiltonianDgDe}) with $%
v_{e}=0$. The atom-field system starts in the state
\begin{equation}
\rho _{AF}(0)=|1,0,g\rangle \langle 1,0,g|
\end{equation}
and evolves to
\begin{equation}
\rho _{AF}(\tau _{A,B})=|c_{1}(\tau _{AB})|^{2}|1,0,g\rangle \langle
1,0,g|+|c_{2}(\tau _{AB})|^{2}|0,1,g\rangle \langle
0,1,g|+(c_{1}(\tau _{AB})c_{2}^{\ast }(\tau _{AB})|1,0,g\rangle
\langle 0,1,g|+h.c.).
\end{equation}
Now the system performs a $\pi $ Rabi pulse, regarded as
instantaneous,
\begin{equation}
\rho _{AF}(\tau _{A,B})=|c_{1}(\tau _{AB})|^{2}|1,0,g\rangle \langle
1,0,g|+|c_{2}(\tau _{AB})|^{2}|0,0,e\rangle \langle
0,0,e|+(c_{1}(\tau _{AB})c_{2}^{\ast }(\tau _{AB})|1,0,g\rangle
\langle 0,0,e|+h.c.).
\end{equation}
Next the atom interacts with $D_{g}$ during a time interval $\tau
_{g}$,
\begin{eqnarray*}
\rho _{AF}(\tau _{A,B}+\tau _{g}) &=&|c_{1}(\tau
_{AB})|^{2}|1,0\rangle \langle 1,0|\left( \int d\mu \langle \psi^{g}
_{\mu }|g\rangle e^{-i\epsilon^{g}_{\mu }\tau _{g}}|\psi^{g}_{\mu
}\rangle \right) \left( \int d\mu \langle g|\psi^{g}_{\mu }\rangle
e^{i\epsilon^{g}_{\mu }\tau _{g}}\langle \psi^{g}_{\mu
}|\right) \\
&+&(c_{1}(\tau _{AB})c_{2}^{\ast }(\tau _{AB})|1,0\rangle \left(
\int d\mu \langle \psi^{g}_{\mu }|g\rangle e^{-i\epsilon^{g}_{\mu
}\tau _{g}}|\psi^{g}_{\mu }\rangle \right) \langle 0,0,e|+h.c.)
\\&+&|c_{2}(\tau _{AB})|^{2}|0,0,e\rangle \langle 0,0,e|,
\end{eqnarray*}
where $\{|\psi^{g}_{\mu }\rangle\}$ and $\epsilon^{g}_{\mu }$
correspond to the set of eigenvectors and eigenvalues of $H_{D}$
with $v_{e}=0$. This atom-detector interaction time will be
considered to have the same \emph{order magnitude} of the $\pi $
Rabi pulse time, and will be disregarded: $\tau _{A,B}+\tau
_{g}\simeq \tau _{A,B}$\footnote{%
On the other hand, it is easy to see that, since the atom and the
fields are not interacting after $t=\tau _{A,B}$, taking the value
of $\tau _{g}$ into account will have no effect on the following
probabilities in this section.}. A click in $D_{g}$ means the atom
was ionized, \emph{i.e.}, its state is described by the set $\left\{
|k\rangle \right\} $; hence the probability of such a click is given
by
\begin{eqnarray}
p_{clickD_{g}} &=&\int dkTr\left\{ |k\rangle \langle k|\rho
_{AF}(\tau
_{A,B}+\tau _{g})\right\} \\
&=&|c_{1}(\tau _{AB})|^{2}p_{g},
\end{eqnarray}
where $p_{g}$ is the efficiency of the detector $D_{g}$:
\begin{equation}
p_{g}=\int dk\left| \int d\mu \langle \psi _{\mu }|g\rangle |\langle
k|\psi _{\mu }\rangle e^{-i\epsilon _{\mu }\tau _{g}}\right| ^{2}.
\end{equation}
If one observes a click in $D_{g}$, the state of the cavity field
collapses to
\begin{equation*}
\rho _{A}(\tau _{A,B})=|1,0\rangle \langle 1,0|,
\end{equation*}
returning to its initial state; thus the probability that $D_{g}$
clicks for the $N$ atoms is
\begin{equation}
P_{clickD_{g}}^{(N)}=\left( |c_{1}(\tau _{AB})|^{2}p_{g}\right)
^{N}. \label{pclickN1}
\end{equation}

Of course in the limit $p_{g}=1$ one recovers the result of the
previous section:
\begin{equation*}
\lim_{N\rightarrow \infty }P_{clickD_{g}}^{(N)}=1.
\end{equation*}
The effect of having an inefficient measurement, \emph{i.e.}, having
a detection efficiency $p_{g}<1$, will change this scenario. This is
illustrated in Fig.1, where we plot the probability of $N$
consecutive clicks in $D_{g}$ as a function of $N$ for different values of $%
p_{g}$. In this case the limit $N\rightarrow \infty $ yields
\begin{equation}
\lim_{N\rightarrow \infty }P_{clickD_{g}}^{(N)}=0.
\end{equation}
This does not mean that Zeno effect is not present. Given the
detector's inefficiency one can not associate the effect to the
statistics of $D_{g}$ clicks: no click in $D_{g}$ does not
necessarily mean that the photon in fact decayed from cavity $A$ to
$B$. The intrinsic detection inefficiency limits the experimental
visibility of the Zeno effect in the present experimental scheme.

\begin{figure}
    \centering
    \includegraphics[scale=0.75]{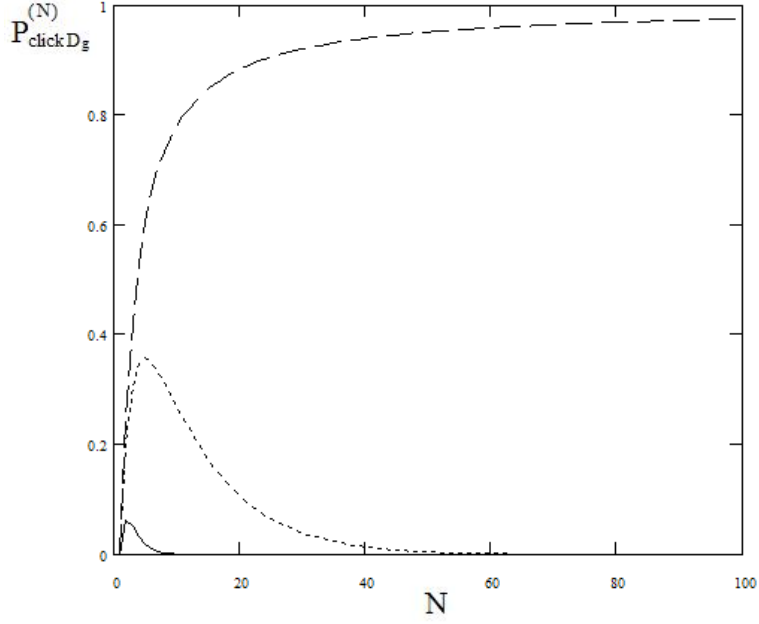}
    \caption{Probability of consecutive clicks in
$D_{g}$ as a function of $N$, for $T=\frac{\pi}{2g}$ and different
values of $p_{g}$: $p_{g}=1$ (dashed), $p_{g}=0,9$ (dotted) and
$p_{g}=0,5$ (continuous)}
    \label{fig1}
\end{figure}

\subsubsection{Only Detector $D_{e}$ is Present}

Another possibility of investigating the limited detection
efficiency in the same experimental scheme consists in having only
detector $D_{e}$ present. This corresponds to the Hamiltonian
(\ref{HamiltonianDgDe}) with $v_{g}=0$.

Note that in this case one click in $D_{e}$ projects the cavity state to $%
|0,0\rangle \langle 0,0|$; thus, in order to observe the effect we
must study sequences of events which do not give rise to any click
in $D_{e}$. In the first step of such a sequence the initial
atom-field state is given by
\begin{equation}
\rho _{AF}(0)=|1,0,g\rangle \langle 1,0,g|,
\end{equation}
which evolves to
\begin{equation}
\rho _{AF}(\tau _{A,B})=|c_{1}(\tau _{AB})|^{2}|1,0,g\rangle \langle
1,0,g|+|c_{2}(\tau _{AB})|^{2}|0,1,g\rangle \langle
0,1,g|+(c_{1}(\tau _{AB})c_{2}^{\ast }(\tau _{AB})|1,0,g\rangle
\langle 0,1,g|+h.c.),
\end{equation}
and next performs an instantaneous $\pi $ Rabi pulse, leading to the
state
\begin{equation}
\bar{\rho}_{AF}(\tau _{A,B})=|c_{1}(\tau _{AB})|^{2}|1,0,g\rangle
\langle 1,0,g|+|c_{2}(\tau _{AB})|^{2}|0,0,e\rangle \langle
0,0,e|+(c_{1}(\tau _{AB})c_{2}^{\ast }(\tau _{AB})|1,0,g\rangle
\langle 0,0,e|+h.c.).
\end{equation}
In the sequence the atom interacts with the detector according to
Eq. (\ref {HamiltonianDgDe}) with $v_{g}=0$, leading to the state
\begin{eqnarray*}
\rho _{AF}(\tau _{A,B}) &=&|c_{2}(\tau _{AB})|^{2}|0,0\rangle
\langle 0,0|\left( \int d\mu \langle \psi^{e}_{\mu }|e\rangle
e^{-i\epsilon^{e}_{\mu }\tau _{e}}|\psi^{e}_{\mu }\rangle \right)
\left( \int d\mu \langle e|\psi^{e} _{\mu
}\rangle e^{i\epsilon^{e}_{\mu }\tau _{e}}\langle \psi^{e}_{\mu }|\right) \\
&&+|c_{1}(\tau _{AB})|^{2}|1,0,g\rangle \langle
1,0,g|+(c_{2}c_{1}^{\ast }|0,0\rangle \left( \int d\mu \langle
\psi^{e}_{\mu }|e\rangle e^{-i\epsilon^{e}_{\mu }\tau
_{e}}|\psi^{e}_{\mu }\rangle \right) \langle 1,0,g|+h.c.),
\end{eqnarray*}
where $\{|\psi^{e}_{\mu }\rangle\}$ and $\epsilon^{e}_{\mu }$
correspond to the set of eigenvectors and eigenvalues of $H_{D}$
with $v_{g}=0$. $\tau _{e}$ is the atom-detector interaction time
which will be neglected as in the previous section. If no click in
$D_{e}$ is observed, the state of the cavity field ends up in
\begin{equation*}
\rho _{F}(\tau _{A,B})=\frac{|c_{1}(\tau _{A,B})|^{2}|1,0\rangle
\langle 1,0|+|c_{2}(\tau _{AB})|^{2}(1-p_{e})|0,0\rangle \langle
0,0|}{|c_{1}(\tau _{AB})|^{2}+|c_{2}(\tau _{AB})|^{2}(1-p_{e})}.
\end{equation*}

This statistical mixture is the initial state of the next step,
whose final state can be calculated in an analogous way as above,
giving
\begin{equation}
\rho _{F}(2\tau _{A,B})=\frac{(|c_{1}(\tau
_{AB})|^{2})^{2}|1,0\rangle \langle 1,0|+|c_{2}(\tau
_{AB})|^{2}(1-p_{e})(1+|c_{1}(\tau _{AB})|^{2})|0,0,\rangle \langle
0,0|}{(|c_{1}(\tau
_{AB})|^{2})^{2}+|c_{2}(\tau _{AB})|^{2}(1-p_{e})(1+|c_{1}(\tau _{AB})|^{2})}%
.
\end{equation}
All subsequent steps will present distinct final states, but always
statistical mixtures of $|1,0\rangle \langle 1,0|$ and $|0,0\rangle
\langle 0,0|$. Since the part related to $|0,0\rangle \langle 0,0|$
does not vary with time, only the part concerning $|1,0\rangle
\langle 1,0|$ will be responsible for changes in the state, which
will be the same in every step, and may be expressed as
\begin{equation}
|1,0\rangle \langle 1,0|\longrightarrow |c_{1}(\tau
_{AB})|^{2}|1,0\rangle \langle 1,0|+|c_{2}(\tau
_{AB})|^{2}(1-p_{e})|0,0\rangle \langle 0,0|. \label{substi}
\end{equation}
Consequently, it is easy to obtain the state of the fields after $i$
no clicks in $D_{e}$:
\begin{equation}
\rho _{F}^{\left( i\right) }\left( i\tau _{A,B}\right)
=\frac{(|c_{1}(\tau _{AB})|^{2})^{i}|1,0\rangle \langle
1,0|+|c_{2}(\tau _{AB})|^{2}(1-p_{e})(\sum_{k=1}^{i}|c_{1}(\tau
_{AB})|^{k-1})|0,0\rangle \langle 0,0|}{(|c_{1}(\tau
_{AB})|^{2})^{i}+|c_{2}(\tau
_{AB})|^{2}(1-p_{e})(\sum_{k=1}^{i}|c_{1}(\tau _{AB})|^{k-1})}.
\end{equation}
The probability of no click in $D_{e}$ in the $i$-th step may be
calculated as
\begin{equation}
p_{\tilde{n}clickD_{e}}^{(i)}=\int dkTr\left\{ (|g\rangle \langle
g|+|e\rangle \langle e|)\rho _{F,A}^{(i)}\right\}
\end{equation}

Where $\rho _{F,A}^{(i)}$ is the state of the system at the N-th
step immediately before the interaction between atom and detector.
This state operator can be calculated from $\rho _{F}^{(i-1)}$.  The
probability of $N$ consecutive no clicks in $D_{e}$ may be computed
as the product
\begin{equation}
P_{\tilde{n}clickD_{e}}^{(N)}=\prod_{i=1}^{N}p_{\tilde{n}clickD_{e}}^{(i)}=
(|c_{1}(\tau _{AB})|^{2})^{N}+|c_{2}(\tau
_{AB})|^{2}(1-p_{e})(\sum_{k=1}^{N}|c_{1}(\tau _{AB})|^{k-1}),
\end{equation}
where $p_{e}$, the efficiency of the detector, is given by
\begin{equation*}
p_{e}=\int dk\left| \int d\mu \langle \psi _{\mu }|e\rangle |\langle
k|\psi _{\mu }\rangle e^{-i\epsilon _{\mu }\tau _{e}}\right| ^{2}.
\end{equation*}
In the limit $N\rightarrow \infty $,

\begin{equation*}
\lim_{N\rightarrow \infty }P_{\tilde{n}clickD_{e}}^{(N)}=1.
\end{equation*}

\begin{figure}
    \centering
    \includegraphics[scale=0.75]{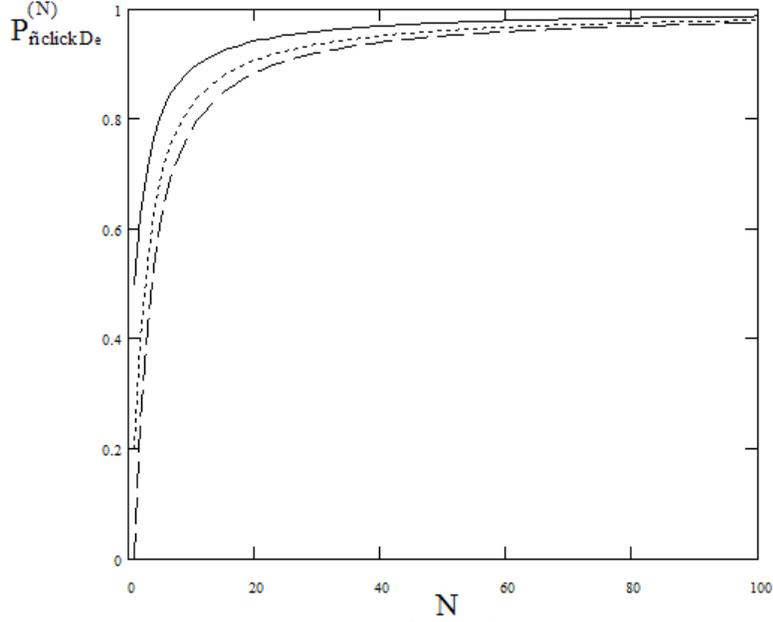}
    \caption{Probability of consecutive no-clicks in $D_{e}$ as a
function of $N$, for $T=\frac{\pi}{2g}$ and different values of
$p_{e}$: $p_{e}=1$ (dashed), $p_{e}=0,8$ (dotted) and $p_{e}=0,5$
(continuous)}
    \label{fig2}
\end{figure}

In Fig.2 we show the probability of $N$ consecutive no-clicks in $%
D_{e}$ for different values of $p_{e}$. For $p_{e}=1$ the curve is
the same as the one for $p_{g}=1$, since no clicks in a perfect
$D_{e}$ is equivalent to clicks in a perfect $D_{g}$. For
inefficient detectors, the probability of $N$ consecutive no-clicks
must be larger than this probability for perfect detectors. This is
illustrated in Fig.2, where the curves representing smaller $p_{e}$
tend to reach the asymptotic value $1$ faster
as $N\rightarrow \infty $. Note that for inefficient detectors no-click in $%
D_{e}$ does not necessarily mean that the photon is for sure in
cavity $A$: the monitoring of the photon transition is not perfect.
However, the asymptotic behavior of $P_{\tilde{n}clickD_{e}}^{(N)}$,
tending to $1$ for any value of $p_{e}$, is most certainly a
consequence of the Zeno effect.

\section{No Intermediate Measurements}

In the experimental set ups discussed in the previous sections the
photon transition was monitored by $N$ probe atoms and a macroscopic
signal was generated. We were interested in the probability of
occurrence of selected sequences, namely, $N$ consecutive clicks in
$D_{g}$ or $N$ consecutive no-clicks in $D_{e}$, which would be
associated to the permanence of the photon in cavity $A$. Obviously,
a complete correlation can not be achieved due to the inefficiency
of the detectors.

Pascazio and Namiki propose in Ref. \cite{art8} a dynamical approach
to QZE and show the essential role of the \emph{generalized spectral decomposition}%
. They propose that QZE occurs even in the absence of intermediate
measures, what explains Itano results in \cite{art7}. For the system
composed by two coupled cavity modes, the generalized spectral
decomposition is brought about by the interaction between the two
level probe atom and the cavity $B$ mode. As we will see, the
classical signals generated by the ionization detectors in each step
(intermediate measures) are not necessary for inhibiting the photon
transition and, accordingly with the approach in \cite{art8}, are
not essential for the characterization of the QZE.

The idea now is to send atoms through cavity $B$, also in $T_{0}/N$
intervals, and not to measure the outcome of the atom-cavity
interaction each time. After $N$ such interactions one atom is sent
through cavity $A$ and measured by a detector $D_{e}$.

As in the previous schemes, the first step of the evolution starts
with the atom-fields state given by
\begin{equation}
\rho _{AF}(0)=|1,0,g\rangle \langle 1,0,g|,
\end{equation}
which evolves to
\begin{equation}
\rho _{AF}(\tau _{A,B})=|c_{1}(\tau _{AB})|^{2}|1,0,g\rangle \langle
1,0,g|+|c_{2}(\tau _{AB})|^{2}|0,1,g\rangle \langle
0,1,g|+(c_{1}(\tau _{AB})c_{2}^{\ast }(\tau _{AB})|1,0,g\rangle
\langle 0,1,g|+h.c.),
\end{equation}
and then to
\begin{equation}
\rho _{AF}(\tau _{A,B})=|c_{1}(\tau _{AB})|^{2}|1,0,g\rangle \langle
1,0,g|+|c_{2}(\tau _{AB})|^{2}|0,0,e\rangle \langle
0,0,e|+(c_{1}(\tau _{AB})c_{2}^{\ast }(\tau _{AB})|1,0,g\rangle
\langle 0,0,e|+h.c.).
\end{equation}
Since this atom is not measured, the field state must be represented
in the end of the step by
\begin{eqnarray}
\rho _{F}(\tau _{AB}) &=&Tr_{A}\left\{ \rho _{AF}(\tau _{A,B})\right\} \\
&=&|c_{1}(\tau _{AB})|^{2}|1,0\rangle \langle 1,0|+|c_{2}(\tau
_{AB})|^{2}|0,0\rangle \langle 0,0|,
\end{eqnarray}
where $Tr_{A}$ is the trace over the variables of the atom, and
accounts for the lack of information about the atomic state.

In order to calculate the final state of the following steps, we
must observe that only the part of $\rho _{A}$ related to
$|1,0\rangle \langle 1,0|$ changes with time, in a way that may be
described by
\begin{equation}
|1,0\rangle \langle 1,0|\longrightarrow |c_{1}(\tau
_{AB})|^{2}|1,0\rangle \langle 1,0|+|c_{2}(\tau
_{AB})|^{2}|0,0\rangle \langle 0,0|.  \label{sub2}
\end{equation}
Thus, it is easy to see that the state operator for the fields in
the cavities, after the interaction of $M_{B}$ with $N$ atoms, can
be written as
\begin{equation}
\rho _{F}(T_{0})=(|c_{1}(\tau _{AB})|^{2})^{N}|1,0\rangle \langle
1,0|+|c_{2}(\tau _{AB})|^{2}\sum_{k=1}^{N}(|c_{1}(\tau
_{AB})|^{2})^{k-1}|0,0\rangle \langle 0,0|.
\end{equation}
The probability that the photon transition from cavity $A$ to cavity
$B$ has not occurred is
\begin{equation}
p_{1,0}^{(N)}=(|c_{1}(\tau _{AB})|^{2})^{N},
\end{equation}
and, in the limit $N\rightarrow \infty $,
\begin{equation}
\lim_{N\rightarrow \infty }p_{1,0}^{(N)}=1.
\end{equation}
This, according to the dynamical approach in \cite{art8},
characterizes Zeno effect. The measurement of this probability can
be done by using one probe atom prepared in $|g\rangle $ state and
sent through cavity $A$ immediately after the interaction of $M_{B}$
with the $N$-th atom. If this probe atom and mode $M_{A}$ perform a
$\pi $ Rabi pulse, the atom-fields state will be given by
\begin{equation}
\rho _{AF}(T_{0})=(|c_{1}(\tau _{AB})|^{2})^{N}|0,0,e\rangle \langle
0,0,e|+|c_{2}(\tau _{AB})|^{2}\sum_{k=1}^{N}(|c_{1}(\tau
_{AB})|^{2})^{k-1}|0,0,g\rangle \langle 0,0,g|,
\end{equation}
and measuring the energy level of the atom with an ionization
detector tell us about the field state. The inefficiency of the
detector enters just as a multiplicative factor in the data.

\subsection{Finite Interaction Times and Lossy Cavities}

The problems related to the inefficiency of the ionization
detectors, which imposed important limitations for the observation
of Zeno effect in the proposals of Sec. III, have been overcome by
the experimental proposal of the present section. However there are
other limitations if a realistic experiment is to be performed.
Firstly the cavity is not perfect and dissipation/decoherence will
also affect the visibility of the effect. And secondly the
interaction time is finite. We consider all these effects in the
present section.

Fig.3 sketches the time evolution, divided in $N$ steps, each one
composed by two parts: no atom is present and the cavities are
coupled (clear zones), the atom interacts with mode $M_{B}$ during a $%
\pi $ Rabi pulse (dark zones). Each clear zone corresponds to the
time interval $\tau _{AB}=T_{0}/N$, where $T_{0}$ is, as in previous
sections, the time during which a photon passes from cavity $A$ to
cavity $B$ if no atom is present: $T_{0}=\pi /2g$. Since our goal
here is to study the inhibition (due to intermediate interactions)
of such a photon transition, the cavities will be uncoupled during
the atom-field interactions, in order to keep the total interaction
time between modes $M_{A}$ and $M_{B}$ fixed
in $T_{0}$ \footnote{%
We must put a piezoeletric element acting in the waveguide, in order
to interrupt the cavities coupling during atom-field interaction.}.
For the Rubydium atoms used in the experiment \cite{art18}, the $\pi
$ Rabi pulse time is $\tau _{\pi }\simeq 10^{-5}s$, and the increase
in the number of probe atoms, $N$, may turn the total time of
atom-field interactions, $N\tau _{\pi }$, quantitatively important.
In order to take this time into account, we must consider
\begin{equation*}
T_{0}^{^{\prime }}=T_{0}+N\tau _{\pi }
\end{equation*}
as the total time of one experimental sequence.

\begin{figure}
    \centering
    \includegraphics[scale=0.7]{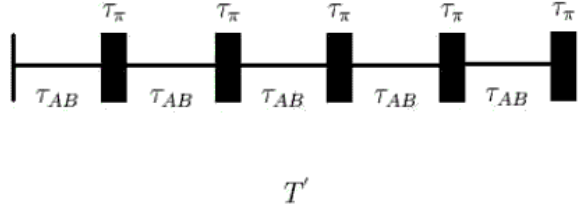}
    \caption{Sketch of the total time of one
experimental sequence.}
    \label{fig3}
\end{figure}

Let us start by modeling the environment as a large set of harmonic
oscillators linearly coupled to the system of interest (modes $M_{A}$ and $%
M_{B}$) \cite{art19}. This model has been used to calculate the time
evolution of two microwave modes constructed in a single cavity, and
the theoretical results showed good agreement with experimental ones
\cite{art20}. In Ref. \cite{art21} it is shown that, for identical
cavities and zero temperature, the model leads to the master
equation
\begin{eqnarray}
\frac{d}{dt}\rho _{F}(t) &=&k\left( 2a\rho _{F}(t)a^{\dagger }-\rho
_{F}(t)a^{\dagger }a-a^{\dagger }a\rho _{F}(t)\right) -i\omega
\left[
a^{\dagger }a,\rho _{F}(t)\right]   \label{evCCdiss} \\
&&+k\left( 2b\rho _{F}(t)b^{\dagger }-\rho _{F}(t)b^{\dagger
}b-b^{\dagger }b\rho _{F}(t)\right) -i\omega \left[ b^{\dagger
}b,\rho _{F}(t)\right]
\notag \\
&&-ig\left[ b^{\dagger }a+a^{\dagger }b,\rho _{F}(t)\right] ,
\notag
\end{eqnarray}
where $\omega $ is the frequency of the modes of interest, $g$ is
their coupling constant and $k$ gives the decay rate of the
cavities; cross decay rates and shifts in $\omega $ and $g$, which
tend to be small \cite{art23}, were disregarded. Using this master
equation, we calculate the time evolution of the state
\begin{equation}
\rho _{F}(0)=|1_{A},0_{B}\rangle \langle 1_{A},0_{B}|=|1,0\rangle
\langle 1,0|  \label{rhoc0}
\end{equation}
as
\begin{equation}
\rho _{F}(t)=(f_{1}(t)|1,0\rangle +l_{2}(t)|0,1\rangle
)(h.c.)+(1-|f_{1}(t)|^{2}-|l_{2}(t)|^{2})|0,0\rangle \langle 0,0|,
\label{rhoc1}
\end{equation}
where
\begin{eqnarray}
f_{1}\left( t\right)  &=&\exp \left[ -\left( k+i\omega \right)
t\right]
\cosh \left[ -igt\right] , \\
l_{2}\left( t\right)  &=&\exp \left[ -\left( k+i\omega \right)
t\right] \sinh \left[ -igt\right] .  \notag
\end{eqnarray}
The probability of finding the photon in cavity $A$, in this case,
is given by
\begin{equation}
|f_{1}(t)|^{2}=e^{-2kt}\cos ^{2}(gt).  \label{f11}
\end{equation}

If the field state has evolved from $t=0$ to $t=$ $\tau _{AB}$ in
the manner
described above, and at time $t=$ $\tau _{AB}$ an atom prepared in $%
|g\rangle $ state begins its interaction with mode $M_{B}$, the
state of the whole system will be given by
\begin{equation}
\rho _{AF}(\tau _{AB})=(f_{1}(\tau _{AB})|1,0,g\rangle +l_{2}(\tau
_{AB})|0,1,g\rangle )(h.c.)+(1-|f_{1}(\tau _{AB})|^{2}-|l_{2}(\tau
_{AB})|^{2})|0,0,g\rangle \langle 0,0,g|.  \label{rhoca2}
\end{equation}
During the atom-field interaction, the field modes evolve
independently, since they are uncoupled. The evolution of state
(\ref{rhoca2}) is described by the master equation
\begin{eqnarray}
\frac{d}{dt}\rho _{AF}(t) &=&k\left( 2a\rho _{AF}(t)a^{\mathbf{\dagger }%
}-\rho _{AF}(t)a^{\mathbf{\dagger }}a-a^{\mathbf{\dagger }}a\rho
_{AF}(t)\right) +i\omega \left[ a^{\mathbf{\dagger }}a,\rho
_{AF}(t)\right]
\label{evJCdiss} \\
&&+k\left( 2b\rho _{AF}(t)b^{\mathbf{\dagger }}-\rho _{AF}(t)b^{\mathbf{%
\dagger }}b-b^{\mathbf{\dagger }}b\rho _{AF}(t)\right) -i\frac{\Omega _{0}}{2%
}[b^{\mathbf{\dagger }}\sigma _{-}+b\sigma _{+},\rho _{AF}(t)],
\notag
\end{eqnarray}
where $\Omega _{0}$ is vacuum Rabi frequency, and $\sigma _{-}=\sigma _{+}^{%
\mathbf{\dagger }}=|g\rangle \langle e|$. The first line of \ Eq.
(\ref {evJCdiss}) describes the dissipation of mode $M_{A}$; the
second line describes the interaction of the atom with mode $M_{B}$
according to the dissipative Jaynes-Cummings model \cite{art23}. In
previous calculations, $\tau _{\pi }$ was the time spent by an atom
to absorb the excitation of mode $M_{B}$. Here, $\tau _{\pi }$ plays
an analogous role, and will be defined as
\begin{equation}
\tau _{\pi }=\frac{1}{\sqrt{\Omega _{0}^{2}-k^{2}}}\arccos \left( \frac{%
2k^{2}-\Omega _{0}^{2}}{\Omega _{0}^{2}}\right) .
\end{equation}
This time, which depends not only on the vacuum Rabi frequency, but
also on the dissipation constant, is the time for a complete
transfer of the excitation of mode $M_{B}$, to the atom or to the
environment. This
definition coincides with the previous one if no dissipation is considered ($%
k=0$). Using master equation (\ref{evJCdiss}) to describe the
evolution of the system from $t=\tau _{AB}$ to $t=$ $\tau _{AB}+\tau
_{\pi }$, we get

\begin{eqnarray}
\rho _{AF}(\tau _{AB}+\tau _{\pi }) &=&|f_{1}(\tau
_{AB})|^{2}e^{-2k\tau _{\pi }}|1,0,g\rangle \langle
1,0,g|+|l_{2}(\tau _{AB})|^{2}e^{-k\tau _{\pi
}}|0,0,e\rangle \langle 0,0,e|  \label{rhoca3} \\
&+&(1-|f_{1}(\tau _{AB})|^{2}e^{-2k\tau _{\pi }}-|l_{2}(\tau
_{AB})|^{2}e^{-k\tau _{\pi }})|0,0,g\rangle \langle 0,0,g|.  \notag
\end{eqnarray}

The state of the fields after the interaction with the first atom is
obtained by taking the trace over the atomic variables:
\begin{eqnarray*}
\rho _{F}(\tau _{AB}+\tau _{\pi }) &=&Tr_{A}\left\{ \rho _{AF}(\tau
_{AB}+\tau _{\pi })\right\} \\
&=&|f_{1}(\tau _{AB})|^{2}e^{-2k\tau _{\pi }}|1,0\rangle \langle
1,0|+(1-|f_{1}(\tau _{AB})|^{2}e^{-2k\tau _{\pi }})|0,0\rangle
\langle 0,0|.
\end{eqnarray*}
Observing that the part of the density operator associated to
$|0,0\rangle \langle 0,0|$ does not change with time, it is easy to
calculate the probability to find the photon in cavity $A$ after the
interaction with $N$ atoms:
\begin{eqnarray}
p_{1,0}^{(N)} &=&(|f_{1}(\tau _{AB})|^{2}e^{-2k\tau _{\pi }})^{N}
\label{probfin} \\
&=&e^{-2k\left( T_{0}+N\tau _{\pi }\right) }\left( \cos ^{2}\left( \frac{%
gT_{0}}{N}\right) \right) ^{N}.  \notag
\end{eqnarray}
This equation explicitates the effect of $N$ intermediate
interactions over
two kinds of temporal dependencies. The term $\left( \cos ^{2}\left( \frac{%
gT_{0}}{N}\right) \right) ^{N}$ represents no transition of the
photon from
cavity $A$ to cavity $B$. It grows when $N$ increases, tending to $1$ when $%
N\rightarrow \infty $. The term $e^{-2k\left( T_{0}+N\tau _{\pi
}\right) }$, related to the probability that the photon has not
decayed to the environment, decreases to zero when $N\rightarrow
\infty $. Of course this decrease is due to the enhancement of the
total time in which the field is exposed to the environment, not
being related to any kind of anti-Zeno effect. Since the dynamics of
dissipation is exponential, it is not affected
by intermediate measures. The role played by the finite interaction time $%
\tau _{\pi }$ is also explicitated and will become quantitatively
important as $N\rightarrow \infty $.

In order to observe the dependence of $p_{1,0}^{(N)}$ on $N$, an
atom prepared in $|g\rangle $ state is sent into cavity $A$ just
after the
interaction of the $N$-th atom with mode $M_{B}$. The atom then performs a $%
\pi $ Rabi pulse, and passes through a $D_{e}$ detector. If the
efficiency of $D_{e}$ is $p_{e}$, the probability of a click will be
given by
\begin{equation}
p_{D_{e}click}^{(N)}=p_{e}e^{-2k\left( T_{0}+N\tau _{\pi }\right)
}\left( \cos ^{2}\left( \frac{gT_{0}}{N}\right) \right) ^{N}.
\end{equation}
This is the empirical quantity to be measured in the present
proposal.

There will be no problems associated to the efficiency $p_{e}$,
since it enters just as a multiplicative factor that does not depend
on $N$. However, the term $e^{-2k\left( T_{0}+N\tau _{\pi }\right)
}$ \emph{depends} on $N$, and may prevent the observation of Zeno
effect if the decay constant $k$ is not small enough. In Fig.4, we
may observe the competition between the tendencies of
$p_{D_{e}click}^{(N)}$ when $N$ grows: the increasing one, due to
Zeno effect, and the decreasing one, due to dissipation. In the
continuous curve $k=10^{3}s^{-1}$, corresponding to the cavities
used in several experiments \cite{art16}. In this case it would be
very difficult to observe Zeno effect, since dissipation dominates
even for small values of $N$. For the dashed curve $k=10s^{-1}$;
this value corresponds to the cavity described in
\cite{art13,art14}, and turns the observation of Zeno effect
possible.

\bigskip

\begin{figure}
    \centering
    \includegraphics[scale=0.73]{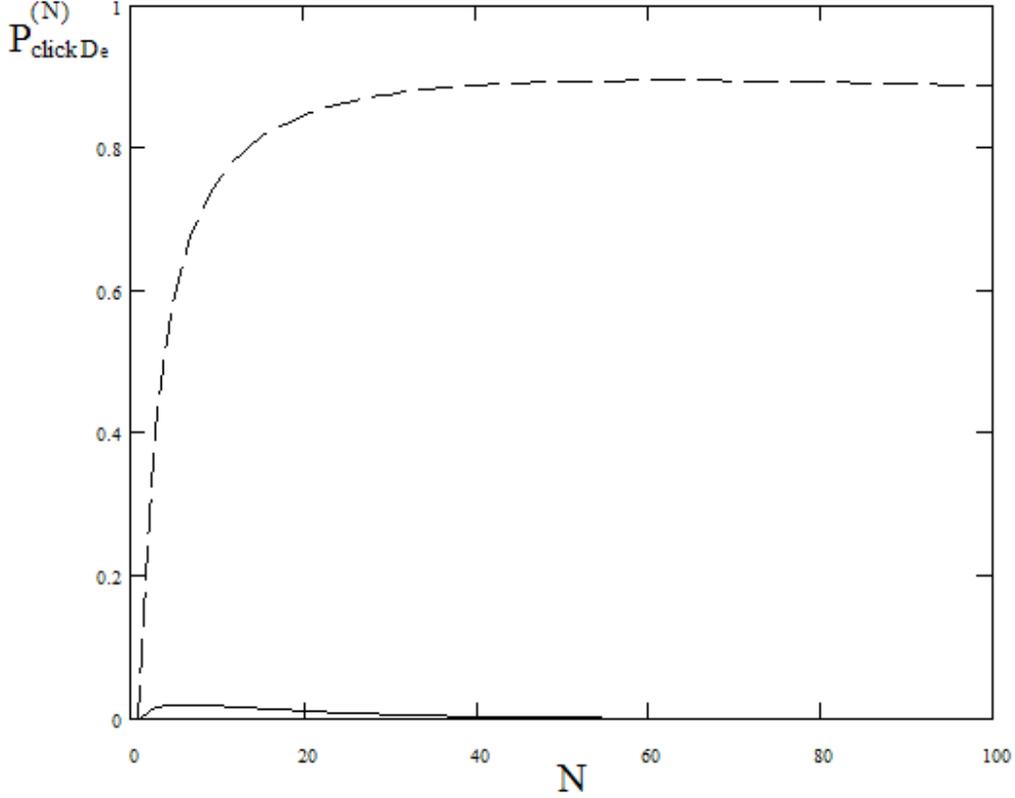}
    \caption{Probability of one click in $D_{e}$
as a function of $N$, for $T=\frac{\pi}{2g}$,
$\Omega_{0}=10^{5}s^{-1}$, $p_{e}=1$, $g=10^{3}s^{-1}$ and different
values of $k$: $k=10^{3}s^{-1}$(continuous) and
$k=10s^{-1}$(dashed).}
    \label{fig4}
\end{figure}

\section{Conclusion}

We consider some realistic aspects related to the observation of the
QZE in Cavity QED. They are: the effect of a lossy environment, of
limited detection efficiency and finite atom-field interaction time.
The calculations are fully analytical and the experiment is
apparently feasible \cite{art13,art14}. Our main result is the
equation for the probability of a no-click detection as a function
of the number of incoming atoms, the cavities dissipation constants,
the probability of a click in detector $D_{e} $ and a finite
atom-field interaction time $\tau _{\pi }$,
\begin{equation}
p_{D_{e}click}^{(N)}=p_{e}e^{-2k\left( T_{0}+N\tau _{\pi }\right)
}\left( \cos ^{2}\left( \frac{gT_{0}}{N}\right) \right) ^{N}.
\end{equation}
This is the main result of the present contribution. It
explicitates, within the context of the present model the role
played on the visibility of the QZE by a realistic apparatus and
realistic detectors. We hope this result may encourage the
experimental realization of the present proposal.

\end{document}